\documentclass[twocolumn,prl,floatfix,a4paper]{revtex4}
\usepackage{blindtext}
\usepackage{comment}
\usepackage[english]{babel}
\usepackage[utf8]{inputenc}
\usepackage[T1]{fontenc}
\usepackage{cancel,xcolor}
\usepackage{caption}
\usepackage{dsfont}
\usepackage{amsmath}
\usepackage{amssymb}
\usepackage{amsfonts}
\usepackage{amsthm}
\usepackage{graphicx}
\usepackage[scale=0.8]{geometry}
\usepackage{float}
\newcommand{\dd}{{\rm d}}

\definecolor{dark-green}{RGB}{0, 128, 0} 

\begin{document}
\title{Unidirectional modes induced by nontraditional Coriolis force in stratified fluids}
\author{Nicolas Perez}
\email{nicolas.perez@ens-lyon.fr}
\author{Pierre Delplace}
\author{Antoine Venaille}
\affiliation{Univ Lyon, ENS de Lyon, Univ Claude Bernard, CNRS, Laboratoire de Physique (UMR CNRS 5672), F-69342 Lyon, France}

\begin{abstract}

Using topology, we unveil the existence of new unidirectional modes in compressible rotating stratified fluids. We relate their emergence to the breaking of time-reversal symmetry by rotation and vertical mirror symmetry by stratification and gravity. We stress the role of the Coriolis force's nontraditional part, induced by a rotation field tangent to the surface. In contrast with horizontally trapped equatorial waves induced by the traditional component of Coriolis force perpendicular to the surface, we find vertically trapped modes that are trapped along interfaces between regions with distinct stratification properties. 
We show that such modes are generalized atmospheric Lamb waves whose direction of propagation can be selected by the nontraditional component of the Coriolis force.

\end{abstract}

\maketitle

\begin{figure}[H]
\begin{center}
\includegraphics[scale=0.25]{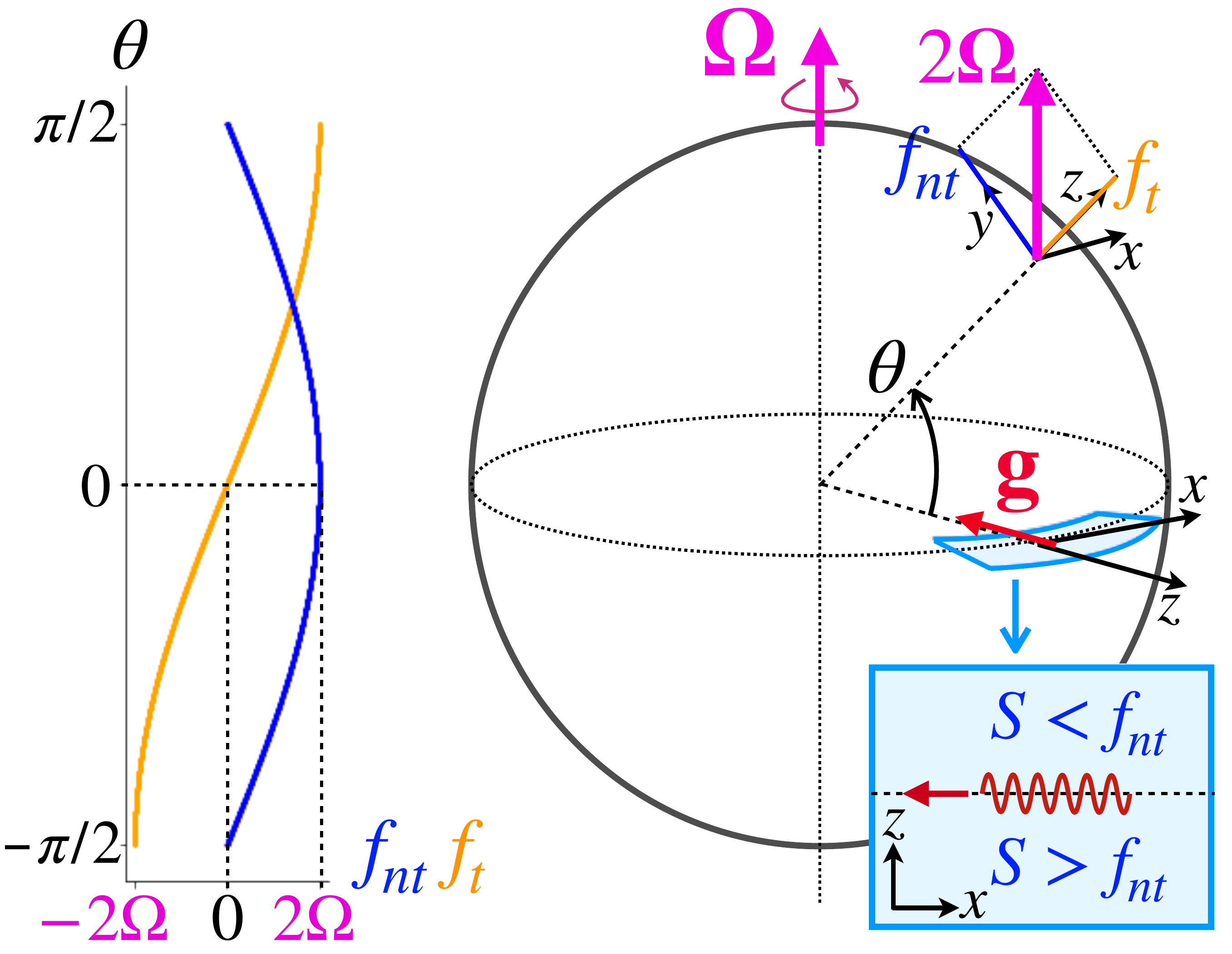}
\end{center}
\caption{\label{fig:Geometry} Geometry of the planar problem on a rotating planet. The traditional component $f_t$ cancels at the equator, while the nontraditional term $f_{nt}$ is maximal. Modes are trapped in the direction $z$ where $S = + f_{nt}$ (or $S = -f_{nt}$) and propagate along the zonal direction $x$.}
\end{figure}

\textbf{Introduction --} When breaking time-reversal symmetry, two-dimensional systems may host unidirectional trapped modes. This is the case in the quantum Hall effect \cite{buttiker1988absence}, plasmas \cite{parker2020topological,fu2021topological}, and certain photonic crystals \cite{photonicReview},  when a magnetic field is applied. Such a symmetry breaking can also be achieved by taking advantage of rotation effects, either at a local level through self-propelled particles \cite{shankar2020topological}, gyroscope arrays \cite{nash2015topological} or curved photonic waveguides \cite{ozawa2019topological}, or at a global scale by a background solid-body rotation through the effect of the Coriolis force \cite{delplace2017topological,wang2015coriolis}. Over planetary scales, atmospheres and oceans are rotating quasi-two-dimensional fluids, and as such they support the propagation of various trapped modes in the horizontal plane \cite{iga1995transition,iga2001transition,godin2012incompressible}. For such flows, it is customary to retain only the component of the Coriolis force that involves the projection of the rotation vector on the local vertical axis, perpendicular to the fluid, which is encoded by the Coriolis parameter $f_t$ (see Fig. \ref{fig:Geometry}). Neglecting the horizontal contribution $f_{nt}$ of the planetary rotation is referred to as the \textit{traditional} approximation \cite{vallis2017atmospheric}. While exact for a truly two-dimensional fluid, this traditional approximation cannot be justified from first principles in more comprehensive three-dimensional geophysical flows. Thereby, the consideration of nontraditional effects has been subjected to a resurgence of interest for the last years \cite{gerkema2008geophysical,tort2014consistent}. Here we report a striking effect due to the nontraditional contribution of the Coriolis force: When combined with compressibility and density stratification, it induces previously unnoticed unidirectional trapped waves in the fluid.

To reveal the existence of such waves, we apply the machinery of topology. Topology offers an elegant theoretical framework to unveil the existence of robust trapped waves along boundaries or interfaces, from quantum to classical wave systems, without having to compute the detailed spectrum of the inhomogeneous system for a specific geometry and set of parameters. In the geophysical realm, this approach shed new light on various well-known waves such as equatorial waves \cite{matsuno1966quasi,delplace2017topological}, coastal Kelvin waves \cite{delplace2020topology} and Lamb waves \cite{lamb1911atmospheric,perrot2019topological}. In the equatorial case, the interface that traps unidirectional waves is defined by the change of sign of the traditional component of the Coriolis parameter $f_t$ at the equator. In contrast, we focus here on a flow taking place in the vertical plane at the equator (Fig. \ref{fig:Geometry}), where this contribution vanishes, so that the only effect of rotation occurs through the nontraditional component $f_{nt}$. This contribution being  constant in that plane, an interface thus cannot be obtained by rotation effects alone. Instead, we show that an emergent interface defined by the change of sign of a stratification parameter involving density gradients, gravity and nontraditional Coriolis force, separates two regions that behave as if they were topologically distinct, thus hosting an unidirectional trapped mode.

\textbf{Flow model --} To describe the consequence of time-reversal symmetry breaking by nontraditional Coriolis effects, we introduce the dynamical equations for adiabatic perturbations of a compressible fluid initially at rest in a uniformly rotating frame at rotation rate $\mathbf{\Omega}$, say a planet's oceanic or atmospheric layer, or a stellar core. We consider propagation phenomena of small length scales compared to the typical radius of the planet or star, and define local Cartesian coordinates (Fig. \ref{fig:Geometry}). The medium is assumed to be inviscid and subjected to a constant gravity $-g \hat{\mathbf{e}}_z$ along the local vertical direction.

The global equilibrium state in the rotating frame has a vertical stratification of density (and temperature, pressure) which is characterized by the buoyancy (or Brunt-Väisälä) frequency $N(z)$ \cite{vallis2017atmospheric}, which is real and positive as we assume a stably stratified medium. The local speed of sound waves, denoted $c_s$, is also assumed to be a function of  $z$ only. To highlight most salient features induced by nontraditional effect of the Coriolis force, we focus on the dynamics taking place in the vertical equatorial plane $(x,z)$ (Fig. \ref{fig:Geometry}), where the traditional component $f_t$ vanishes. The dynamics is then restricted to a two-dimensional flow in the $(x,z)$ plane.

We consider the wave dynamics governed by the  primitive equations -- conservation of momentum, mass and entropy -- linearized around the state of rest. By performing a suitable rescaling of the dynamical fields with the density profile, those equations can be recast in a Schr\"odinger-like matricial form as ${\rm i} \partial_t \psi = \hat{\mathcal{H}} [z,\partial_x,\partial_z] \psi$, with a four-component vector field $\psi = \begin{pmatrix} u & w & \Theta & p \end{pmatrix}^t$, where $u \hat{\mathbf{e}}_{x} + w \hat{\mathbf{e}}_{z}$ is the rescaled perturbation's velocity field in the normalized basis $(\hat{\mathbf{e}}_{x}, \hat{\mathbf{e}}_{z})$, $\Theta$ and $p$ are the rescaled perturbed potential temperature and pressure respectively (see Supplemental Material). The wave operator reads
\begin{equation} \label{eq:operator}
    \begin{split}
        &\hat{\mathcal{H}} =\\
        &\begin{pmatrix} 0 & -{\rm i} f_{nt} & 0 & -{\rm i} c_s \partial_x \\ {\rm i} f_{nt} & 0 & -{\rm i} N & {\rm i} S - \frac{{\rm i} c_s '}{2} - {\rm i} c_s \partial_z \\ 0 & {\rm i} N & 0 & 0 \\ -{\rm i} c_s \partial_x & -{\rm i} S - \frac{{\rm i} c_s '}{2} - {\rm i} c_s \partial_z & 0 & 0 \end{pmatrix} \ ,
    \end{split}
\end{equation}
where $c_s '$ stands for the derivative of $c_s$ along the vertical coordinate $z$, and the stratification parameter $S(z)$ is defined as \cite{perrot2019topological}:
\begin{equation} \label{eq:stratification_parameter}
    S(z) = \frac{c_s}{2 g} \left( N^2 - \frac{g^2}{c_{s}^2} \right) - \frac{c_s '}{2} \ .
\end{equation}
This stratification parameter contains all the physical ingredients breaking the system's mirror symmetry in the vertical direction $z$ (gravity $g$, stratification of density through $N$, and non-uniformity of $c_s$), while time-reversal symmetry is broken only by the constant nontraditional Coriolis parameter $f_{nt}$.

\begin{figure}[t]
\begin{center}
\includegraphics[scale=0.42]{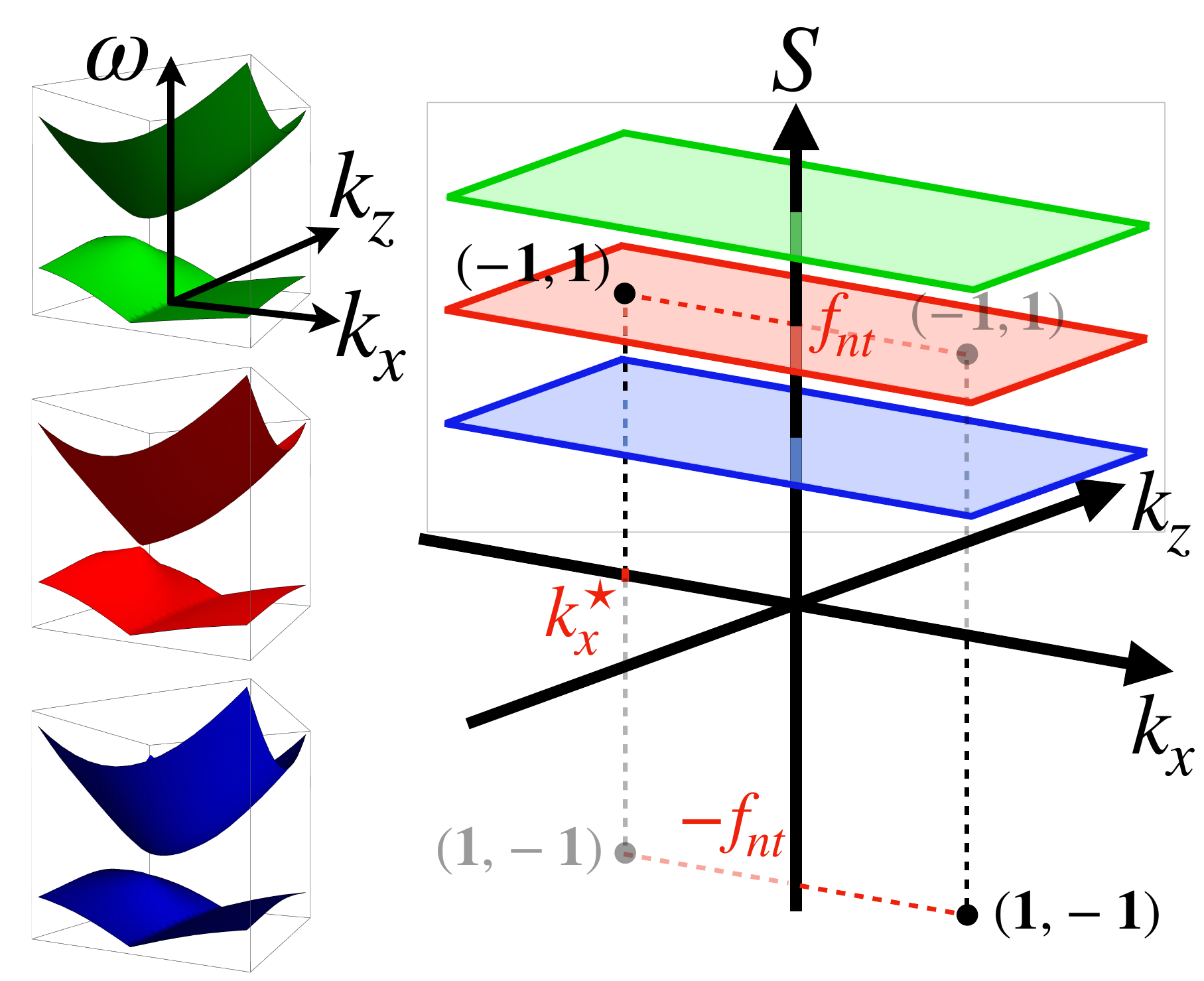}
\end{center}
\caption{\label{fig:Chern} Degeneracy points (in black) between the positive-frequency bands ($n=+1$ and $+2$) of the spectrum of \eqref{eq:symbol}, with $f_{nt} > 0$. The pairs of integers indicate the topological charges $(\mathcal{C}_1,\mathcal{C}_2)$ of the two bands involved in a given degeneracy point. The insets on the left show their dispersion relations associated with the green, red and blue planes in parameter space. We also show the degeneracy points (grey) between the negative-frequency bands (not shown on the left), with $(\mathcal{C}_{-2},\mathcal{C}_{-1})$.}
\end{figure}

\textbf{Topological properties of the bulk waves --} The physical model of interest is inhomogeneous in the $z$ direction, through $N(z)$, $S(z)$ and $c_s(z)$. However, important spectral properties of the operator \eqref{eq:operator}, and in particular the existence of unidirectional trapped modes, can be inferred from the topology of the eigenmodes $\Psi_n$ of the much simpler eigenvalue problem $\omega_n \Psi_n = H \Psi_n$ with
\begin{equation} \label{eq:symbol}
    H = \begin{pmatrix} 0 & - {\rm i} f_{nt} & 0 & c_s k_x \\ {\rm i} f_{nt} & 0 & - {\rm i} N & {\rm i} S + c_s k_z  \\ 0 & {\rm i} N & 0 & 0 \\ c_s k_x & - {\rm i} S+c_s k_z  & 0 & 0 \end{pmatrix} \ .
\end{equation}
This dual ''bulk'' problem describes plane-wave solutions $\psi(x,z,t) = \Psi_n {\rm e}^{{\rm i} \left( k_x x + k_z z - \omega t \right)}$ of \eqref{eq:operator}, valid in the case of an unbounded $(x,z)$ plane, with constant parameters $N$, $S$ and $c_s$. In mathematical terms, the matrix $H$ is called the \textit{symbol} of the operator $\hat{\mathcal{H}}$, and is bijectively associated to it through the Wigner-Weyl transform \cite{moyal1949quantum} (see Supplemental Material). In the general situation (where the plane is bounded and the parameters are functions of $z$), plane waves are not solutions of the problem, however the symbol matrix \eqref{eq:symbol} can still be formally defined as a function of $z$ (through the parameters $N$, $S$ and $c_s$) and the wave vector $\mathbf{k} = k_x \hat{\mathbf{e}}_{x} + k_z \hat{\mathbf{e}}_{z}$. For a given triplet $(z,k_x,k_z)$, the symbol matrix $H$ admits four eigenvalues $\omega_n$, corresponding to two low-frequency internal-gravity wavebands ($n = \pm 1$) and two high-frequency acoustic wavebands ($n = \pm 2$), both modified by rotation (see Fig. \ref{fig:Chern}). Because of the realness of the fields, each eigenvalue $\omega$ of $H$ at $(z,k_x,k_z)$ is associated with the eigenvalue $-\omega$ at $(z,-k_x,-k_z)$. Therefore, it is sufficient to consider positive frequencies ($n = 1, 2$) in the following.

Building upon previous topological analysis in geophysics \cite{delplace2017topological,perrot2019topological,delplace2020topology}, we expect the existence of unidirectional trapped modes in the inhomogeneous problem \eqref{eq:operator} to be ruled by topological indices of the symbol's eigenmodes $\Psi_n$. More specifically, these \textit{topological} modes are expected to arise around a typical wavenumber $k_x^\star$, and to be trapped around a height $z^\star$, provided wavebands of the symbol cross at an isolated point $(z^\star,k_x^\star,k_z^\star)$ in the $(z,k_x,k_z)$ space. In that case, it is possible to assign topological charges to such degeneracy points, given by topological indices called the Chern numbers $\mathcal{C}_n$, that account for the twisting of the eigenmode bundles around the band-crossing point in $(z,k_x,k_z)$ space (see Supplemental Material). According to index theorems \cite{NakaharaBook}, we expect a number of $\mathcal{C}_n$ trapped modes whose frequency branches transit towards the waveband $n$ of the inhomogeneous problem when swapping $k_x$.

The first step of our strategy then consists in identifying the degeneracy points of the symbol's eigenvalues. We find such crossing points between the gravity ($n = 1$) and the acoustic ($n = 2$) wavebands precisely when $S = + f_{nt}$ and $S = - f_{nt}$ (Fig. \ref{fig:Chern}), respectively designated with the $\pm$ sign in the following paragraphs. Those are located at $\left( z^\star , k_x^\star , k_z^\star \right)$, with $k_z^\star = 0$, $k_x^\star = \mp \sqrt{f_{nt}^{2} + N(z^\star)^2} / c_{s}(z^\star)$ and $z^\star$ such that $S(z^\star) = \pm f_{nt}$: their existence thus results from the combination of mirror symmetry breaking (due to $S$) and time-reversal symmetry breaking (due to $f_{nt}$). Since the degeneracy condition depends on $z$ through $S$ only, one can consider the more convenient parameter space of $(S,k_x,k_z)$ to compute the Chern numbers of the symbol (see Supplemental Material).

Once the degeneracy points are found out, one needs to compute the Chern numbers  associated to the eigenstate bundles involved in the waveband crossing. This is a standard calculation that can always be performed at least numerically, by means of the integration of a local quantity, called the Berry curvature \cite{berry1984quantal}, over a closed surface enclosing the degeneracy point in $(S,k_x,k_z)$ space (see Supplemental Material). Actually, since this degeneracy simply involves a generic conical two-band crossing point, the possible values of the Chern numbers of the two eigenstate bundles are known to be opposite and of absolute value $1$ (see e.g. the appendix of \cite{delplace2020topology}). Therefore, one can directly infer the existence of one unidirectional wave, trapped around the interface line $z=z^\star$ in the vertical $(x,z)$ plane, that is where the quantity $S(z)\mp f_{nt}$ changes sign.

\begin{figure*}[t]
\begin{center}
\includegraphics[scale=0.43]{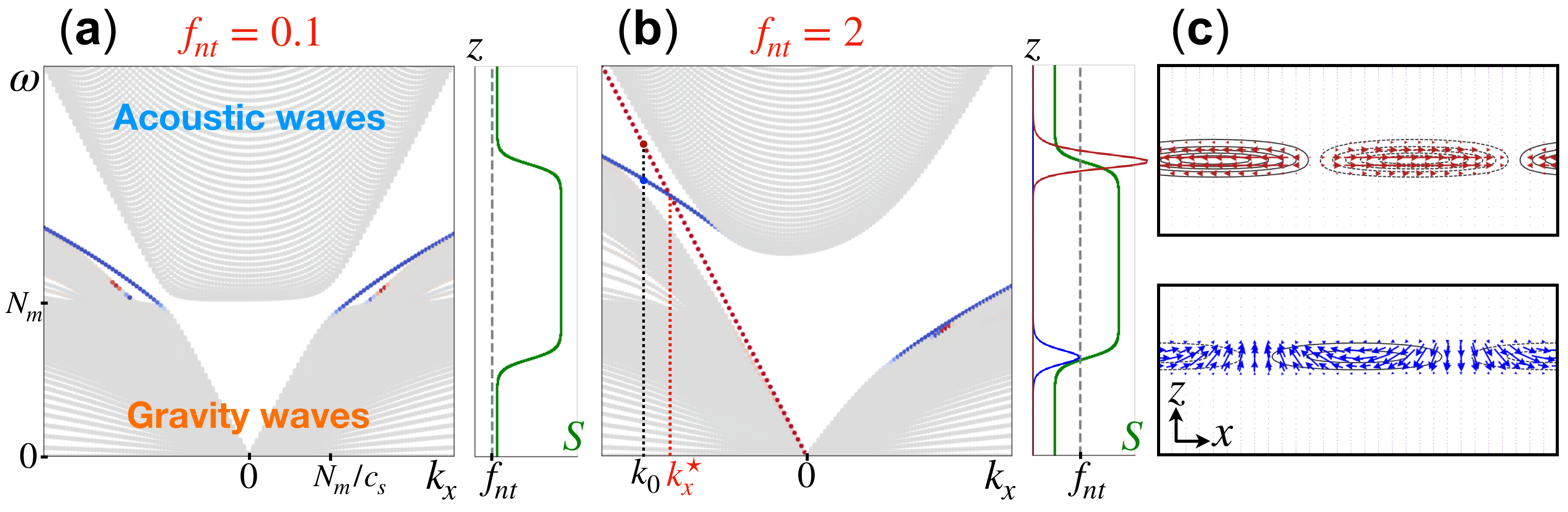}
\end{center}
\caption{\label{fig:Spectra} Numerical solutions of the eigenvalue problem \eqref{eq:operator} projected on the Fourier basis -- such that $(\partial_x,\partial_t) \rightarrow {\rm i}(k_x,-\omega)$ --, with constant sound speed $c_s$, obtained with Dedalus \cite{burns2020dedalus}. ($\bold{a}$) Spectrum for $f_{nt}=0.1$: $S$ does not take the value $f_{nt}$, no mode transits through the gap. $N_m = \min \{ N(z) \}$. ($\bold{b}$) $f_{nt} = 2$: $S$ takes twice the value of $f_{nt}$, a mode is localised at each interface $S = f_{nt}$ and transits through the gap around $k_x^\star$. The pressure perturbations's amplitude $|\Tilde{p}(z)|$ of the topological modes at $k_x = k_0$ are plotted with the background stratification $S$ in green. ($\bold{c}$) Plot of the velocity field of the topological modes with wavenumber $k_0$, with pressure levels in background (positive values in solid lines and negative in dashed lines).}
\end{figure*}

\textbf{Unidirectional modes induced by the nontraditional Coriolis component --}
We then check this prediction numerically by computing the frequency spectrum of $\hat{\mathcal{H}}$ for solutions of the form $\psi(x,z,t) = \begin{pmatrix} \Tilde{u}(z) & \Tilde{w}(z) & \Tilde{\Theta}(z) & \Tilde{p}(z) \end{pmatrix}^t {\rm e}^{{\rm i} (k_{x} x - \omega t)}$, with varying parameters in the vertical direction $z$. We can always consider $f_{nt} > 0$, since the eastward (increasing $x$) and westward (decreasing $x$) directions can be defined with the direction of planetary rotation. For numerical convenience, we also consider the profile of $S(z)>0$ depicted in Fig. \ref{fig:Spectra}, that is periodic in $z$ and crosses the value $f_{nt}$ in both an increasing and a decreasing fashion (the degeneracy point involved in this numerical simulation is thus the one at $S = +f_{nt}$, but a general conclusion is given in the next paragraph, including for the case $S = -f_{nt}$), so that we can discuss, in the same simulation, the two opposite slopes $S'(z^\star)>0$ and $<0$ at an interface where $S(z)-f_{nt}$ changes sign. In the case where $S(z)$ does not reach the value of the nontraditional Coriolis parameter $f_{nt}$, (Fig. \ref{fig:Spectra}(a)), there is no mode that transits between the gravity and the acoustic waveband, as expected from the topological analysis, since there is no degeneracy point of the symbol and thereby no topological charge. In contrast, if the profile of $S(z)$ is such that $S(z)-f_{nt}$ changes sign (Fig. \ref{fig:Spectra}(b)), a topological charge arises in the symbol picture at $(z^\star, k_x^\star,0)$, and accordingly, a new mode appears in the spectrum of $\hat{\mathcal{H}}$ around $k_x=k_x^\star$. This mode transits from the internal-gravity waveband to the acoustic one, so that the number of modes gained or lost by the two wavebands when sweeping $k_x$ is correctly given by the Chern numbers $\pm1$. Such a \textit{spectral flow} from one waveband to another implies the sign of the group velocity of the corresponding wave to be fixed, hence its unidirectionality. We have also computed the spreading of this mode in the $z$ direction and found that it is indeed localized where $S - f_{nt}$ changes sign (in color in Fig. \ref{fig:Spectra}).

The numerical analysis reveals interesting differences on the nature of the topological trapped modes depending on the sign of the slope of $S(z)$ at the interface where $S(z) -f_{nt}$ changes sign. The corresponding rescaled velocity fields are shown in Fig. \ref{fig:Spectra}(c), and their typical vertical trapping length around the interface is $\mathcal{L}^\star = \sqrt{c_{s}(z^\star)/|S'(z^\star)|}$ \cite{perrot2019topological}: the steeper the slope of $S$ at the crossing point, the more localised the modes.

\begin{itemize}
    \item If $S$ decreases, a first unidirectional mode appears (red in Fig. \ref{fig:Spectra}). For an interface $S = +f_{nt}$ this mode propagates westward, lies around $k_x^\star = -\sqrt{f_{nt}^2 + N(z^\star)^2}/c_{s}(z^\star)$ and transits from the acoustic waveband to the gravity waveband (as $k_x$ increases around the value $k_x^\star$) in the positive-frequency spectrum. For $S = -f_{nt}$ it propagates eastward, lies around $k_x^\star = \sqrt{f_{nt}^2 + N(z^\star)^2}/c_{s}(z^\star)$ and transits from the gravity waveband to the acoustic waveband in the positive-frequency spectrum. In the case where $c_s$ is constant (e.g. in an isothermal ideal gas) this mode is longitudinal ($w=0$), non-dispersive, with dispersion relation $\omega = \pm c_s k_x$ if $S(z^\star) = \mp f_{nt}$ (Fig. \ref{fig:Spectra}(b)). It thus shares strong similarities with the Lamb-like wave discussed in \cite{perrot2019topological} and the equatorial Kelvin wave \cite{delplace2017topological}. However its nature is different: while it is unidirectional and confined in the vertical direction, the equatorial Kelvin wave is confined in the meridional direction due to the gradient of $f_t$, the traditional Coriolis parameter -- which vanishes here --, and Lamb-like waves are not unidirectional because their existence does not rely on time-reversal symmetry breaking. Therefore this topological mode can be seen as the rotating counterpart of the Lamb-like waves in stratified compressible fluids. We have checked that this wave still exists when $c_s$ varies with $z$, as expected from our topological analysis, although it becomes dispersive.
    
    \item If $S$ increases, another unidirectional mode of opposite spectral flow (i.e. transiting from the gravity waveband to the acoustic wave band if $S(z^\star) = f_{nt}$ and the other way round if $S(z^\star) = -f_{nt}$) appears around $k_x^\star$ (blue in Fig. \ref{fig:Spectra}). Contrary to the previous topological mode, this one has a non-longitudinal velocity field (Fig. \ref{fig:Spectra}(c)). It is reminiscent of a zero-group-velocity oscillation at frequency $N$, modified by rotation and varying buoyancy frequency $N(z)$. As their group velocities have the same sign, both modes propagate in the same direction, that is westward if $S(z^\star) = +f_{nt}$ and eastward if $S(z^\star) = -f_{nt}$.
\end{itemize}

According to this analysis, a topological transition occurs when the number of crossings between the vertical profile $S(z)$ and the value $\pm f_{nt}$ changes, leading to the appearance/disappearance of unidirectional interface waves whose frequency branches transit between the gravity and acoustic wavebands. These modes shall not be confused with the ones described by Iga \cite{iga1995transition,iga2001transition}, whose existence depend on the boundary conditions, neither with those addressed by Godin \cite{godin2012incompressible} who describes an incompressible wave motion solution of the compressible dynamics with a free surface and traditional rotation, whereas compressibility, nontraditional rotation and internal interfaces are key ingredients of the modes studied in this letter.

\textbf{Discussion --} Figure \ref{fig:Spectra} depicts an ideal situation where $S$ takes the value $+f_{nt}$ and not $-f_{nt}$. Westward topological modes are thus allowed to propagate whereas the eastward ones are not. This shows that rotation selects the direction of propagation of the Lamb-like waves studied in \cite{perrot2019topological}. In realistic planetary fluids, $f_{nt}$ is however expected to be much smaller. On Earth, $f_{nt} \approx 10^{-4}$ s$^{-1}$ remains negligible compared to the typical values of $S$ in the atmosphere ($\approx -10^{-2}$ s$^{-1}$), and the trapping length in the ocean is higher than its typical depth (see Supplemental Material of \cite{perrot2019topological}). Therefore the Earth's atmosphere and the oceans are not fitted for potential observations of these topological modes.

Nevertheless, if $S$ changes sign at a given $z_0$, then we expect the topological modes' location to be shifted even if $f_{nt} > 0$ is small: if $S'(z_0) < 0$, the eastward Lamb-like mode should be located at $z > z_0$ and the westward one at $z < z_0$, and conversely if $S'(z_0) > 0$. As far as applications are concerned, this effect of splitting the topological modes' location in the weak rotational regime is a feature that is more likely to be observed, especially in the astrophysical context. In fact, possible observations of Lamb-like waves in asteroseismology (the study of stellar pulsations) have been previously argued, yet without discussing the effect of rotation  \cite{perrot2019topological,leclerc2022topo}. Indeed, using the model of a polytropic star with constant adiabatic index, $S$ is a decreasing function of the radius $r$, that varies from $+\infty$ at the center of the star to $-\infty$ at the surface \cite{leclerc2022topo}. It thus cancels at some radius $r_0$ and generically takes both values $\pm f_{nt}$ for any rotation rate. Therefore, in rotating stellar interiors, one can expect a westward Lamb-like mode closer to the core ($r<r_0$) and an eastward one closer to the surface ($r>r_0$), thus improving the chance of a possible direct observation of the latter. These modes are well-separated if $f_{nt}$ is high enough so that the topological modes do not completely overlap, in other words if the sum of their respective extension in the radial direction, $\mathcal{L} = \sqrt{c_{s}/|S'|}_{S=\pm f_{nt}}$, is not high compared to the distance $d$ separating the interfaces. This could be achieved for some young stars of the upper main sequence of the Hertzprung-Russell diagram, which can reach high rotation rates. For instance, Vega, in the Lyra constellation, has a rotation period of about 0.7 days \cite{takeda2008rotational}, thus $f_{nt} \approx 2 \, 10^{-4}$ s$^{-1}$ at the equator, yielding a ratio of order 1 between $d$ and the sum of the trapping lengths.
%
%

\textbf{Conclusion --} Our study shows the essential role of the nontraditional Coriolis force in the propagation properties of rotating  stratified fluid waves. Thanks to topological arguments, we have found new unidirectional waves transiting between the low-frequency gravity and high-frequency acoustic wavebands, that are trapped along lines determined by a balanced between rotation and vertical stratification. 
The existence of these modes is robust to perturbations in the stratification profile until the interface constraint $S = \pm f_{nt}$ is broken. All salient features of the two-dimensional case presented in this letter can be extended to the  three-dimensional case with full Coriolis force, which will be explained in a companion paper. The application of this work to asteroseismology provides an promising line of research for future investigations of topological waves in natural media, in particular in the presence of non-Hermitian effects induced by dissipation, mean flows and convection.

\section{Supplemental Material}

\textbf{Model equations and fields rescaling --} We introduce a local basis $(x,y,z)$, where $z$ is normal to the sphere, so that the gravity be $\mathbf{g} = -g \hat{\mathbf{e}}_z$, $x$ follow the longitudinal direction pointing eastward and $y$ the meridional direction pointing northward (Fig. \ref{fig:Geometry}). In that local basis the solid body rotation rate has two components and we note
\begin{equation} \label{eq:Coriolis_vector}
    2 \mathbf{\Omega} = \begin{pmatrix} 0 \\ f_{nt} \\ f_t \\ \end{pmatrix} = 2 \Omega \begin{pmatrix} 0 \\ \cos{\theta} \\ \sin{\theta} \\ \end{pmatrix} \ ,
\end{equation}
where $\Omega$ is the value of the rotation rate and $\theta$ the latitude. We are interested in small perturbations of an inviscid fluid initially at rest in the rotating frame. Absorbing the centrifugal force into the gravity $\mathbf{g}$, Euler's equation reads
\begin{equation} \label{eq:NS}
    \frac{\partial \mathbf{v}}{\partial t} + (\mathbf{v} \cdot \nabla) \mathbf{v} = -\frac{1}{\rho} \nabla P - g \hat{\mathbf{e}}_z - 2 \mathbf{\Omega} \times \mathbf{v} \ ,
\end{equation}
where $P$ is the pressure, $\rho$ the density and $\mathbf{v}$ the velocity field.

We also introduce the equation of conservation of mass,
\begin{equation} \label{eq:mass_conservation}
    \frac{\partial \rho}{\partial t} + \nabla \cdot (\rho \mathbf{v}) = 0 \ ,
\end{equation}
and the equation of conservation of entropy (assuming adiabatic evolution of fluid parcels),
\begin{equation} \label{eq:entropy_conservation}
    \frac{\partial \rho}{\partial t} + (\mathbf{v} \cdot \nabla) \rho = c_{s}^{-2} \left[ \frac{\partial P}{\partial t} + (\mathbf{v} \cdot \nabla) P \right] \ .
\end{equation}
The isentropic compressibility factor of the fluid is noted $\chi_s = c_{s}^{-2}$, where $c_s$ is the local celerity of sound waves. For the fluid at rest, we have $\mathbf{v} \equiv \mathbf{0}$ and the fields (noted with a $0$ index) are stationary ($\partial_t \equiv 0$), hence the hydrostatic relation
\begin{equation} \label{eq:static_equilibrium}
    \nabla P_0 = -g \rho_0 \hat{\mathbf{e}}_z \ .
\end{equation}
Therefore the fluid is stratified along the $z$ direction and relation \eqref{eq:static_equilibrium} reads
\begin{equation} \label{eq:static_equilibrium_bis}
    \frac{\dd P_0 (z)}{\dd z} = -g \rho_0 (z) \ .
\end{equation}
We also consider that $c_s$ is a function of $z$. We now write the perturbation fields around the stationary state of rest as
\begin{equation} \label{eq:perturbations}
    \mathbf{v}' \equiv \mathbf{v}, \quad \rho' \equiv \rho - \rho_0, \quad P' \equiv P - P_0 \ ,
\end{equation}
and linearize equations \eqref{eq:NS}, \eqref{eq:mass_conservation} and \eqref{eq:entropy_conservation}:
\begin{subequations} \label{eq:linearized_equations}
    \begin{align}
        \rho_0 &\frac{\partial \mathbf{v}'}{\partial t} = -\nabla P' - g \rho' \hat{\mathbf{e}}_z - 2 \rho_0 \mathbf{\Omega} \times \mathbf{v}' \ , \\
        &\frac{\partial \rho'}{\partial t} + v_{z}'\frac{\dd \rho_0}{\dd z} = -\rho_0 \nabla \cdot \mathbf{v}' \ , \\
        &\frac{\partial \rho'}{\partial t} + v_{z}'\frac{\dd \rho_0}{\dd z} = c_{s}^{-2} \left( \frac{\partial P'}{\partial t} - g \rho_0 v_{z}' \right) \ .
    \end{align}
\end{subequations}
It is now convenient to use the following transformations \cite{perrot2019topological}:
\begin{equation} \label{eq:square_root}
    \mathbf{u} \equiv \rho_{0}^{1/2} \: \mathbf{v}' \equiv (u,v,w), \quad \Tilde{\rho} \equiv \rho_{0}^{-1/2} \: \rho', \quad \Tilde{P} \equiv \rho_{0}^{-1/2} \: P' \ ,
\end{equation}
so that equations \eqref{eq:linearized_equations} can be rewritten as
\begin{subequations} \label{eq:linearized_equations_bis}
    \begin{align}
        &\partial_t \mathbf{u} = -\nabla \Tilde{P} - g \Tilde{\rho} \hat{\mathbf{e}}_z - 2 \mathbf{\Omega} \times \mathbf{u} - \frac{1}{2 \rho_0} \frac{\dd \rho_0}{\dd z} \Tilde{P} \hat{\mathbf{e}}_z \ , \\
        &\partial_t \left( \Tilde{\rho} - c_{s}^{-2} \Tilde{P} \right) = - \left( \frac{1}{\rho_0} \frac{\dd \rho_0}{\dd z} + g c_{s}^{-2} \right) w \ , \\
        &\partial_t \left( \frac{\Tilde{P}}{c_s} \right) = - c_s \nabla \cdot \mathbf{u} + \left( \frac{g}{c_s} + \frac{c_s}{2 \rho_0} \frac{\dd \rho_0}{\dd z} \right) w \ .
    \end{align}
\end{subequations}
At this point we assume that the quantity
\begin{align}
    - \left( \frac{1}{\rho_0} \frac{\dd \rho_0}{\dd z} + g c_{s}^{-2} \right) \ ,
\end{align}
that is a function of $z$, is strictly positive everywhere. This ensures the hyperbolicity of the set of partial differential equations \eqref{eq:linearized_equations_bis}. Physically it means that the fluid is stably stratified. Therefore we can define the following frequencies:
\begin{equation} \label{eq:frequencies}
    \begin{split}
    &N(z) \equiv \sqrt{-\left( \frac{g}{\rho_0} \frac{\dd \rho_0}{\dd z} + \frac{g^2}{c_{s}^2} \right)} \\
    \text{and} \quad &S(z) = \frac{c_s}{2 g} \left( N^2 - \frac{g^2}{c_{s}^2} \right) - \frac{c_s '}{2} \ .
    \end{split}
\end{equation}
The Brunt-Väisälä frequency $N(z)$ \cite{vallis2017atmospheric} is the pulsation at which a fluid particle oscillates vertically in the stably stratified fluid, at position $z$. The derivative of $c_s$ with respect to $z$ is noted $c_s'$.

Finally we define the following fields for perturbations:
\begin{equation} \label{eq:fields}
    p \equiv \frac{\Tilde{P}}{c_s} \quad \text{and} \quad \Theta \equiv \frac{g}{N} (\Tilde{\rho} - c_{s}^{-2} \Tilde{P}) \ .
\end{equation}
The second one corresponds to the perturbation of potential temperature. Using definitions \eqref{eq:frequencies} and \eqref{eq:fields} along with equations \eqref{eq:linearized_equations_bis}, we finally get
\begin{subequations} \label{eq:linearized_equations_final}
    \begin{align}
        &\partial_t \mathbf{u} = - 2 \mathbf{\Omega} \times \mathbf{u} - N \Theta \hat{\mathbf{e}}_z + \left( S - \frac{c_{s}'}{2} \right) p \hat{\mathbf{e}}_z - c_s \nabla p \ , \\
        &\partial_t \Theta = N w \ , \\
        &\partial_t p = - c_s \nabla \cdot \mathbf{u} - \left( S + \frac{c_{s}'}{2} \right) w \ ,
    \end{align}
\end{subequations}
that we can write in matrix form
\begin{equation} \label{eq:linearized_equations_matrix}
    \partial_t \psi = \hat{\mathcal{H}} \psi \ ,
\end{equation}
with $\psi = \begin{pmatrix} u & v & w & \Theta & p \end{pmatrix}^t$ and the operator
\begin{equation} \label{eq:linearized_equations_matrix_bis}
    \begin{split}
    &\hat{\mathcal{H}} = \\
    &\begin{pmatrix} 0 & f_t & -f_{nt} & 0 & - c_s \partial_x \\ -f_t & 0 & 0 & 0 & - c_s \partial_y \\ f_{nt} & 0 & 0 & -N & S - \frac{c_{s}'}{2} - c_s \partial_z \\ 0 & 0 & N & 0 & 0 \\ - c_s \partial_x & - c_s \partial_y & -S - \frac{c_{s}'}{2} - c_s \partial_z & 0 & 0 \end{pmatrix}
    \end{split}
\end{equation}
Among the parameters of expression \eqref{eq:linearized_equations_matrix_bis}, $N$, $S$ and $c_s$ are functions of $z$, whereas $f_t$ and $f_{nt}$ are functions of the meridional coordinate $y$. In many geophysical situations, the planetary radius is much higher than the variation lengthscales in the vertical direction, therefore we can consider $f_t$ and $f_{nt}$ constant when considering fluid motions at much smaller scale than the radius. In order to emphasize the nontraditional effect, we consider in our work the particular case $f_t = 0$ (equatorial region), which allows one to discard the meridional dependence of the problem (i.e. equation \eqref{eq:linearized_equations_matrix} is compatible with $v$, $\partial_y = 0$). The problem thus conveniently reduces to a four-by-four matrical equation with parameters depending on $z$ only. Thus follows the Schr\"odinger-like form ${\rm i} \partial_t \psi = \hat{\mathcal{H}} [z,\partial_x,\partial_z] \psi$, introduced in equation \eqref{eq:operator}.

\textbf{Wavebands and Chern numbers --} We introduce the \textit{symbol} $H[z,k_x,k_z]$ of the operator $\hat{\mathcal{H}} [z,\partial_x,\partial_z]$, defined by the following relation, called the Weyl correspondence \cite{moyal1949quantum,onuki2020quasi}:
\begin{equation} \label{eq:Wigner}
    \begin{split}
    &\hat{\mathcal{H}} [z,\partial_x,\partial_z] \Psi(x,z) = \frac{1}{(2 \pi)^2} \int H \left[ \frac{z + z'}{2}, k_x, k_z \right] \\
    &\times \Psi(x',z') {\rm e}^{{\rm i} [k_x (x-x') + k_z (z-z')]} \: {\rm d} x' {\rm d} z' {\rm d} k_x {\rm d} k_z \ .
    \end{split}
\end{equation}
To clarify formula \eqref{eq:Wigner}, the symbol is a 4-by-4 matrix-valued function and $\hat{\mathcal{H}}$ is an operator applying on test functions $\Psi$ defined over the $(x,z)$ space. For a given $(x,z)$, $\Psi(x,z) \in \mathbb{C}^4$. Moreover, in this case, both $H$ and $\hat{\mathcal{H}}$ are Hermitian for their respective canonical inner products.

Since the scalar operator $-{\rm i} c_s (z) \partial_z$ is associated to the scalar symbol $c_s k_z + \frac{\rm i}{2} \frac{{\rm d} c_s}{{\rm d} z}$, we get
\begin{equation} \label{eq:symbolz}
    \begin{split}
    &H[z,k_x,k_z] = \\
    &\begin{pmatrix} 0 & - {\rm i} f_{nt} & 0 & c_s (z) k_x \\ {\rm i} f_{nt} & 0 & - {\rm i} N(z) & c_s (z) k_z + {\rm i} S(z) \\ 0 & {\rm i} N(z) & 0 & 0 \\ c_s (z) k_x & c_s (z) k_z - {\rm i} S(z) & 0 & 0 \end{pmatrix}
    \end{split}
\end{equation}
When all coefficients $N$, $S$ and $c_s$ are constant, one may naturally look at the plane-wave solutions of the Schr\"odinger-like equation, i.e. $\psi(x,z,t) \propto {\rm e}^{{\rm i} (k_x x + k_z z - \omega t)}$. In this case the dispersion relation is given by the eigenvalues of the symbol $H$, and the polarization relations between the different fields for an eigenmode are given by the corresponding eigenvector of $H$.

Roughly speaking, the index theorems tell us that a singularity in the spectrum of the operator $\hat{\mathcal{H}}$ shall be generated by a corresponding topological singularity of the symbol $H$. While the first may be quite difficult to identify, the second is more straightforward: for a given point $(z,k_x,k_z)$ the symbol $H$ has four eigenvalues (except at degeneracy points), thus defining four wavebands (indexed by $n = -1, -2, 1, 2$) in this space. We can also define four \textit{bundles} with the (normalized) eigenvectors $\Psi_n [z,k_x,k_z]$ defined on the parameter space. A topological singularity point in parameter space may then arise for a waveband whenever the corresponding family of eigenvectors or bundle can't be continuously defined on any closed surface surrounding this point. These singularities correspond to the degeneracy or crossing points between the wavebands, which are much easier to identify. As explained in the main text, a pair of degeneracy points appear in the bulk spectrum when $z$ is such that $S(z) = \pm f_{nt}$: one between bands $1$ and $2$ at $(k_x,k_z) = (\mp \frac{\sqrt{N^2 + f_{nt}^2}}{c_s},0)$ and its symmetric $-\omega$ between bands $-2$ and $-1$ at $(\pm \frac{\sqrt{N^2 + f_{nt}^2}}{c_s},0)$. Since the degeneracy condition involves the variable $z$ only through $S$, we will work for convenience in the parameter space $(S,k_x,k_z)$ instead of $(z,k_x,k_z)$ (this does not change the values of the Chern numbers).

In order to quantify the topological charge associated to a degeneracy point between two bands, we first define the Berry curvature \cite{berry1984quantal} of the waveband $n$ as
\begin{equation} \label{eq:Berry_curvature}
    \mathbf{F}^{(n)}[S,k_x,k_z] = \nabla \times \left( {\rm i} \Psi_{n}^\dagger \nabla \Psi_n \right)
\end{equation}
where $\nabla$ is the gradient in parameter space $(S,k_x,k_z)$. One can check that this quantity is real. In our case the parameter space is three-dimensional, therefore the Berry curvature \eqref{eq:Berry_curvature} can be expressed as a three-dimensional vector. As the external derivative of the $n^{th}$ bundle's connection ${\rm i} \Psi_{n}^\dagger \nabla \Psi_n$, the Berry curvature is a relevant quantity to identify topological singularities. More precisely, one can show that the flux $(1 / 2 \pi) \oint \mathbf{F}^{(n)} \cdot {\rm d} \mathbf{\Sigma}$ through any closed surface in parameter space is an integer (${\rm d} \mathbf{\Sigma}$ pointing in the outward direction). Much like the Gauss formula, this integer counts the degree of a singularity (or multivaluedness) enclosed by the integration surface. Therefore we can associate a topological charge, called a Chern number \cite{berry1984quantal}, to an isolated degeneracy point of the waveband $n$, as
\begin{equation}
    \mathcal{C}_n = \frac{1}{2 \pi} \oint \mathbf{F}^{(n)} \cdot {\rm d} \mathbf{\Sigma} \ ,
\end{equation}
where the integral is performed on a closed surface enclosing only the crossing point considered. As a degeneracy point involves two wavebands (say $n$ and $m$), topological charges appear by pair at this point, with $\mathcal{C}_n = - \mathcal{C}_m$.

In flow models, as explained in the main text, the index theorem manifests through the so-called bulk-boundary correspondence \cite{delplace2017topological,perrot2019topological,delplace2020topology}, relating the topological charge $\mathcal{C}_n$ to the number of modes transiting between the corresponding wavebands of operator $\hat{\mathcal{H}}$ when crossing a degeneracy point.

\bibliographystyle{apsrev4-1}
\bibliography{Bibliography}

\end{document}